\documentclass[a4paper,11pt]{article}
\usepackage{jcappub} 

\usepackage{xspace}
\usepackage{lineno}
\usepackage{xcolor}

\newcommand{\crpropa}{\texttt{CRPropa3.2.1}\xspace}
\newcommand{\photonbg}{$\gamma\text{BG}$\xspace}
\newcommand{\cnb}{C$\nu$B\xspace}

\arxivnumber{2606.00498} 
\title{\boldmath Suppression of boosted relic neutrinos by photon backgrounds during ultra-high-energy cosmic ray propagation}

\author{Gabriel Azeredo}
\author{and Vitor de Souza}
\affiliation{Instituto de Física de São Carlos, Universidade de São Paulo,\\Av. Trabalhador São Carlense, 400, São Carlos, SP, 13566-590, Brazil}

\emailAdd{gabriel@ifsc.usp.br}
\emailAdd{vitor@ifsc.usp.br}

\abstract{Constraining the cosmic neutrino background (\cnb) represents a major experimental challenge in cosmology. Recent studies have suggested that relic neutrinos boosted by ultra-high-energy cosmic rays (UHECRs) may generate observable diffuse neutrino fluxes. Previous estimates have not effectively propagated the primary cosmic rays, often neglecting crucial energy losses and the unavoidable, competing interactions with diffuse photon backgrounds. Here we revisit these expectations using a realistic Monte Carlo propagation framework. This approach allows us to consistently incorporate cosmic ray energy losses, nuclear photodisintegration, and production of secondary neutrinos. We show that interactions with diffuse photon backgrounds strongly suppress the boosted relic neutrino flux predicted in simplified propagation scenarios. Furthermore, we demonstrate that to produce any observable suppression on the UHECR energy spectrum at Earth, or for the boosted \cnb component to become comparable to the cosmogenic neutrino flux, the \cnb density must be enhanced by a factor, the so-called overdensity, of extreme magnitude ($\eta \gtrsim 10^{8}$).}

\begin{document}
\maketitle
\flushbottom

\section{Introduction}
\label{sec:introduction}

Neutrinos are the second most abundant known particles in the Universe. In the standard cosmological scenario, the active neutrino species decoupled from the primordial plasma at temperatures of order MeV, well before photon decoupling and the formation of the cosmic microwave background (CMB). This relic population forms the cosmic neutrino background (\cnb) \cite{yanagisawa_2014_looking_for_cosmic, giunti_2007_fundamentals_of_neutrino}. The present-day number density is expected to be $n_0 \simeq 336~\text{cm}^{-3}$, including neutrinos and antineutrinos of all active species \cite{de_salas_2016_relic_neutrino_decoupling, akita_2020_precision_relic_neutrino_decoupling, baumann_2022_cosmology}. Small non-thermal distortions generated during neutrino decoupling have been studied in precision calculations, but they do not alter the basic conclusion that the relic neutrino background is extremely abundant and very cold \cite{Froustey_2020, de_salas_2016_relic_neutrino_decoupling}. Its direct detection remains experimentally challenging because of the very low laboratory-frame energies of relic neutrinos \cite{ringwald_2009_prospects_for_the_direct, giunti_2007_fundamentals_of_neutrino}. In extended cosmological scenarios, dark matter \cite{mckeen_2019_cosmic_neutrino} or dark energy \cite{berghaus_2021_cosmology_dark_energy, berghaus_2024_cosmology_dark_energy} may generate additional low-energy neutrinos at late times, giving rise to a overdensity. The strongest direct experimental bound on such an overdensity is currently set by KATRIN, $\eta \lesssim 10^{11}$ \cite{katrin_colaboration_2022_new_constraint_on_the}. 

Ultra-high-energy cosmic rays (UHECRs) are predominantly atomic nuclei with energies above $10^{18}~\text{eV}$. Their spectrum and composition are measured mainly by large hybrid air-shower observatories, in particular the Pierre Auger Observatory in the Southern Hemisphere and the Telescope Array in the Northern Hemisphere \cite{pierre_auger_2015_observatory, telescope_array_2008_experiment}. Both experiments have reported precision measurements of the UHECR energy spectrum and its suppression at the highest energies \cite{pierre_auger_2020_energy_spectrum, telescope_array_2023_energy_spectrum}. During extragalactic propagation, UHECR nuclei interact with the CMB and the extragalactic background light through photopair production, photopion production, and photodisintegration, while also losing energy adiabatically because of the expansion of the Universe. These photon-background interactions are central to the interpretation of the observed UHECR spectrum and to the production of cosmogenic neutrinos \cite{greisen_1966_end_to_the_cosmic, zatsepin_kuzmin_1966_upper_limit}.

The same UHECRs can also scatter off relic neutrinos, transferring part of their energy to the \cnb and producing boosted relic neutrinos. This possibility is relevant for high-energy and ultra-high-energy neutrino observatories, including IceCube, IceCube-Gen2, GRAND, POEMMA, and related radio-detection experiments \cite{Aartsen_2017, icecube_2021_icecube_gen2_the_window, grand_2019_the_giant_radio_array, poemma_2021_observatory}. Recent studies have revisited this mechanism in several contexts. For the diffuse boosted-\cnb component, Ref.~\cite{herrera_2025_diffuse_boosted_cosmic} found that, for an overdensity of $\eta=150$, neutrinos up-scattered by protons could reach the projected sensitivity of future experiments. Ref.~\cite{zhang_2026_impact_of_coherent} extended the calculation by including coherent scattering and a mixed UHECR composition, obtaining bounds as strong as $\eta \lesssim 10^{7}$ for $m_\nu=0.1~\text{eV}$. Other works have considered boosted relic-neutrino production in specific astrophysical environments, including blazars and galaxy clusters \cite{demarchi_2025_relic_neutrino_background_cosmic_ray, monsalvatje_2024_upper_limits_on}, motivated in part by high-energy neutrino observations such as the IceCube event associated with TXS~0506+056 \cite{icecube_collaboration_2018_neutrino_emission}.

Most existing calculations of boosted relic neutrinos are based on line-of-sight integrals or simplified prescriptions for the parent UHECR flux. These approaches are useful and physically transparent, but they make it difficult to account self-consistently for all propagation effects that compete with \cnb scattering. In particular, the UHECR population capable of boosting relic neutrinos is simultaneously depleted and reshaped by interactions with photon backgrounds, nuclear photodisintegration, secondary decay chains, and continuous energy losses. In addition, extrapolating the UHECR spectrum measured at Earth back to the extragalactic medium is not unique, because the observed spectrum depends on the source distribution, injection composition, maximum rigidity, source evolution, and propagation losses. These limitations can lead to an overestimate of the boosted-\cnb flux if the primary cosmic rays are not propagated consistently.

In this work, we present a propagation-based Monte Carlo calculation of boosted-\cnb neutrino production during UHECR propagation. We implement neutral-current scattering of relic neutrinos by nucleons and nuclei in \crpropa, a widely used astroparticle-propagation framework \cite{alves_batista_2022_crpropa_3.2}. This allows us to propagate UHECRs and secondary neutrinos consistently while including interactions with photon backgrounds, hereafter denoted \photonbg, nuclear photodisintegration, nuclear decay chains, and continuous energy losses. The same simulation framework therefore provides the neutrino fluxes from three channels: boosted relic neutrinos from \cnb scattering, cosmogenic neutrinos from interactions with \photonbg, and neutrinos from nuclear decays. To our knowledge, this is the first Monte Carlo implementation of relic-neutrino scattering during UHECR propagation within CRPropa.

Because the mechanisms of UHECR acceleration and injection are still uncertain, we first consider idealized pure-proton and pure-iron source compositions in order to identify the physical regimes in which the boosted-\cnb contribution can become relevant. We then study the dependence on the source spectral index and maximum rigidity. Finally, we use the astrophysical model and combined-fit parameters from the Pierre Auger Observatory \cite{pierre_auger_2017_combined_fit_of_spectrum_and_composition} to obtain more realistic predictions for different overdensity factors. Our goal is to reassess previous expectations for boosted relic-neutrino fluxes under a UHECR propagation framework that consistently accounts for competing photon-background interactions and continuous energy losses.

Because neutrino oscillations measure mass-squared differences rather than the absolute mass scale, the relic neutrino background is composed of several mass eigenstates rather than a single species with a unique mass. In this work, $m_\nu$ should therefore be understood as an effective target-neutrino mass used to characterize the scattering kinematics. We adopt $m_\nu=0.1~\mathrm{eV}$ as a representative benchmark, consistent with the mass scale commonly used in previous studies of boosted relic neutrinos, and leave a full treatment of the three mass eigenstates and their separate phase-space distributions for future work.

This paper is organized as follows. In Section~\ref{sec:interactions_in_the_extragalactic_medium}, we describe the relevant nucleus-photon interactions, decay processes, and the two \cnb-scattering channels implemented in \crpropa. Section~\ref{sec:simulations_and_models} presents the simulation setup and the adopted astrophysical source models. Section~\ref{sec:results} gives the resulting UHECR and neutrino fluxes, and Section~\ref{sec:conclusion} summarizes our conclusions.

\section{Interactions in the extragalactic medium}
\label{sec:interactions_in_the_extragalactic_medium}

After leaving their acceleration sites, the most energetic cosmic rays can interact with diffuse target particles in the extragalactic medium. The four fundamental processes that play a major role in UHECR propagation are photopair production, photopion production, photodisintegration, and adiabatic losses caused by the expansion of the Universe. 

The photopair reaction ($A + \gamma\rightarrow A + e^+ + e^-$) is a particular case of Bethe--Heitler pair production in the Coulomb field of a charged cosmic ray \cite{bethe_1934_stopping_fast_particles,1970_blumenthal_energy_loss_of_high}. Since we are not interested in the development of electromagnetic cascades, this interaction acts as a low-inelasticity energy loss mechanism ($K \sim 10^{-3}$) \cite{2009_dermer_high_energy_radiation}. Consequently, it is can be treated as a continuous energy loss in this study. The continuous regime is already implemented in the standard propagation codes~\cite{alves_batista_2022_crpropa_3.2, aloisio_2017_simprop_v2r4}. 

Photodisintegration is the process by which cosmic-ray nuclei interact with the \photonbg and fragment into lighter nuclei and free nucleons. Because this process can produce unstable nuclei, decay chains must also be taken into account during propagation. These interactions can severely alter the mass composition injected by the sources before it is observed at Earth. Distinct physical channels are dominant depending on the energy scale. In the cosmic-ray rest frame, the giant dipole resonance (GDR) dominates at photon energies of $\varepsilon'_\gamma \lesssim 30\text{--}50~\text{MeV}$, while quasi-deuteron (QD) emission takes over at energies of $50 \lesssim \varepsilon'_\gamma \lesssim 150~\text{MeV}$ \cite{2005_khan_photodisintegration, 2019_rachen_interaction_processes}. The precise values of the photodisintegration cross sections remain a significant source of uncertainty in UHECR propagation models \cite{alves_batista_2015_effects_on_uncertainties_in_simulations, pierre_auger_2017_combined_fit_of_spectrum_and_composition, pierre_auger_collaboration_2023_constraining_the_sources_of_ultra}.

Propagation codes also account for adiabatic energy losses due to the expansion of the Universe, which affect ultra-high-energy cosmic rays and neutrinos independently of their energy.

\subsection{Neutrinos from interactions with \texorpdfstring{$\gamma$}{γ
}BG and decays}

Soon after the discovery of the CMB, it was proposed that the production of pions from interactions between protons and these diffuse photons would yield a suppression in the UHECR flux at Earth, the so-called GZK cutoff \cite{greisen_1966_end_to_the_cosmic, zatsepin_kuzmin_1966_upper_limit}. Pions carry $\sim 20\%$ of the parent nucleon's energy and subsequently decay. Neutral pions decay into gamma rays ($\pi^0\rightarrow 2\gamma$), each carrying $\sim 10\%$ of the initial energy, while charged pions decay into neutrinos and positrons ($\pi^+\rightarrow e^+ + \nu_e + \bar{\nu}_\mu + \nu_\mu$), with each lepton taking $\sim 5\%$ of the initial nucleon energy \cite{alves_batista_2019_secondary_neutrino}. 

Additionally, free neutrons produced as secondaries in photodisintegration or photopion production will undergo $\beta$-decay ($n \rightarrow p + e^- + \bar{\nu}_e$). The resulting electron antineutrinos carry $\sim 0.01\text{--}0.04\%$ of the parent neutron energy \cite{taylor_2007_the_propagation, 2009_dermer_high_energy_radiation}. The dominance of the GDR channel for cosmic-ray nuclei makes it difficult for heavy mass compositions injected at the source to produce more neutrinos than a pure-proton injection \cite{chakraborty_2024_a_relook_at_the_gzk}. 

\subsection{Up-scattered C\texorpdfstring{$\nu$}{ν}B neutrinos}

A cosmic-ray nucleus with energy $E_{\text{cr}}$ and mass $m_{\text{cr}}$ propagating through the \cnb can scatter off a low-energy relic neutrino. For a target neutrino of mass $m_{\nu}$, assuming that $E_{\text{cr}} \gg m_{\text{cr}}\gg m_{\nu}$, the maximum transferred energy is \cite{cappiello_2019_reverse_direct_detection, bringmann_2019_novel_direct}
\begin{equation}
E_{\nu,\max}(E_{\text{cr}}) =
\frac{E_{\text{cr}}^2}{E_{\text{cr}} + m_{\text{cr}}^2/(2m_{\nu})}.
\end{equation}
Detailed calculations of the cross sections for both channels considered in this work are provided in Refs. \cite{zhang_2026_impact_of_coherent, demarchi_2025_relic_neutrino_background_cosmic_ray, giunti_2007_fundamentals_of_neutrino}. Our numerical implementation is primarily based on these derivations. In this section, we present the relevant scattering cross sections, focusing on the incoherent and coherent regimes. Furthermore, we detail how these interactions were incorporated into \crpropa.

For a cosmic-ray nucleus with $E_{\text{cr}} \gtrsim 10^{19}~\text{eV}$, the $Z$ boson can resolve individual nucleons within the nucleus. The differential cross section for this process can be approximated as a linear superposition of neutrino-nucleon scattering over all constituent nucleons:
\begin{equation}
\frac{d\sigma_{\text{incoh}}}{dE_{\nu}} = \left[Z\frac{d\sigma^{\nu p}_{\text{ES}}}{dE_{\nu}} + (A - Z)\frac{d\sigma^{\nu n}_{\text{ES}}}{dE_{\nu}}\right](1 - F^2(q^2)),
\end{equation}
where the superscripts $\nu p$ and $\nu n$ denote the differential cross sections for standard neutrino-proton and neutrino-neutron elastic scattering, respectively; $Z$ is the number of protons and $F(q^2)$ is the nuclear form factor \cite{klein_1999_exclusive}. Using the index $N = p, n$ to denote the specific nucleon, the elastic scattering cross section for a UHE cosmic ray scattering off a non-relativistic neutrino is taken from Ref. \cite{zhang_2026_impact_of_coherent}:
\begin{equation}
\frac{d\sigma^{\nu N}_{\text{ES}}}{dE_{\nu}} = \frac{G_F^2 m_\nu m^4_N}{\pi(s - m_N^2)^2}\left[A_N(q^2) + C_N(q^2)\frac{(s - u)^2}{m^4_N}\right],
\end{equation}
where this expression is summed over both neutrino and antineutrino scattering. Here, $s$, $u$, and $q^2$ are the standard Mandelstam variables, and $m_N$ is the nucleon mass. The coefficient functions in terms of the form factors $A_N(q^2)$ and $C_N(q^2)$ were obtained from Ref. \cite{zhang_2026_impact_of_coherent}.

At energies $E_{\text{cr}} \lesssim 10^{19}~\text{eV}$, coherent scattering, where the neutrino scatters off the nucleus as a whole, becomes dominant. The cross section for this process is given by \cite{zhang_2026_impact_of_coherent}
\begin{equation}
\frac{d\sigma_{\text{coh}}}{dE_{\nu}} = \frac{G_F^2m_\nu}{\pi}Q^2_{W,i}\left[1 - \frac{E_\nu}{E_{\text{cr}}} - \frac{m^2_{\text{cr}}E_\nu}{2m_\nu E^2_{\text{cr}}}\right]F^2(q^2),
\end{equation}
\noindent where the expression is again summed over both neutrinos and antineutrinos. The nuclear weak charge is computed as $Q_{W,i} = Zg_V^p + N g_V^n$, with $g_V^p$ and $g_V^n$ being the vector coupling constants for the proton and neutron, respectively. Summing both channels, the total differential cross section for scattering between a non-relativistic neutrino and an ultra-relativistic cosmic ray is written as
\begin{equation}
    \frac{d\sigma}{dE_{\nu}} \equiv \frac{d\sigma_{\text{coh}}}{dE_{\nu}} + \frac{d\sigma_{\text{incoh}}}{dE_{\nu}}.
\end{equation}
The structure of our neutrino scattering implementation was heavily based on the native Compton scattering module of \crpropa, with the necessary modifications to account for non-relativistic target neutrinos instead of background photons. We pre-tabulated the function
\begin{equation}
F(E_{\nu},E_{\text{cr}}) = \int_{E_{\nu,\min}}^{E_{\nu}} d\bar{E}_{\nu}~\frac{d\sigma}{d\bar{E}_{\nu}}(\bar{E}_{\nu}, E_{\text{cr}}),
\end{equation}
\noindent since the total cross section can be expressed as $\sigma(E_{\text{cr}}) = F(E_{\nu, \max}(E_{\text{cr}}), E_{\text{cr}})$, the mean free path is given by
\begin{equation}
\label{eq:fmp_definition}
\lambda_\eta(E_{\text{cr}}) = \frac{1}{\eta n_0\sigma(E_{\text{cr}})}.
\end{equation}
Here, $\eta$ is treated as a phenomenological overdensity factor multiplying the homogeneous cosmological relic-neutrino density $n_0$. This prescription neglects possible spatial inhomogeneities in the relic-neutrino distribution. Equation~\eqref{eq:fmp_definition} defines the mean free path at $z=0$. During propagation, however, the relic-neutrino density evolves with redshift. In the standard cosmological scenario, the number density scales as $n_\nu(z) = n_0(1+z)^3$. Consequently, in our simulations, the interaction tables for each fixed $m_\nu$ are computed at $z=0$ and rescaled during propagation according to $\lambda_\eta(E_{\text{cr}}, z) = (1+z)^{-3}\lambda_\eta(E_{\text{cr}})$.

\begin{figure}[h]
        \begin{minipage}[b]{0.49\linewidth}
            \centering
            \includegraphics[width=\textwidth]{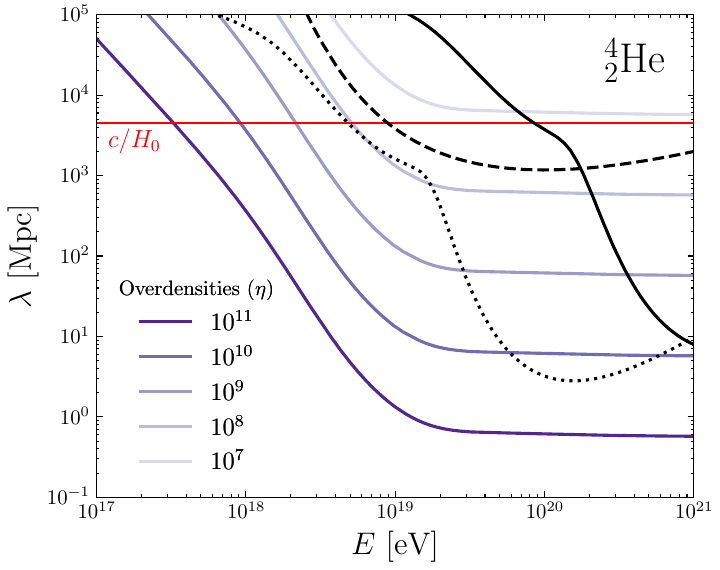}
        \end{minipage}
        \hspace{0.5cm}
        \begin{minipage}[b]{0.49\linewidth}
            \centering
            \includegraphics[width=\textwidth]{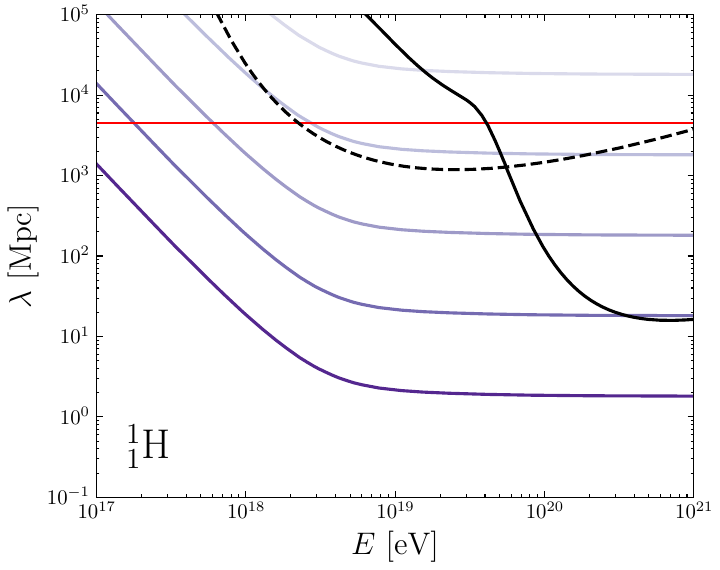}
        \end{minipage} 
        \caption{Mean free paths for the relevant UHECR interactions in the extragalactic medium at $z=0$. The left and right panels display the interactions for Helium-4 ($^4_2\text{He}$) and protons ($^1_1\text{H}$), respectively. Colors denote the mean free path for up-scattered \protect\cnb neutrinos assuming different overdensity factors ($\eta = 10^7\text{--}10^{11}$). Black lines represent interactions with the \photonbg: solid for photopion production, dashed for photopair production, and dotted for photodisintegration. The red line represents adiabatic energy losses due to cosmic expansion.}
        \label{fig:mean_free_path}
\end{figure}

During propagation, the mean free path is used to determine whether an interaction occurs within a given step. If a scattering event occurs, a secondary neutrino is produced and subsequently propagated. The energy of this secondary neutrino is sampled using the cumulative distribution function (CDF),
\begin{equation}
\text{CDF}(E_{\nu},E_{\text{cr}})  = \frac{1}{\sigma(E_{\text{cr}})}\int_{E_{\nu,\min}}^{E_{\nu}} d\bar{E}_{\nu}~\frac{d\sigma}{d\bar{E}_{\nu}}(\bar{E}_{\nu}, E_{\text{cr}}).
\end{equation}
The mean free paths for interactions with diffuse photons and neutrinos at $z = 0$ are shown in Fig.~\ref{fig:mean_free_path}. These results yield important physical insights: for overdensities $\eta \lesssim 10^7$, C$\nu$B scattering is virtually negligible at low redshifts. Conversely, for extreme overdensities $\eta \gtrsim 10^{10}$, interactions with diffuse neutrinos become comparable to those with diffuse photons for primary protons and helium nuclei. This trend can be broadly generalized to heavier nuclei.

\section{Simulation and astrophysical models}
\label{sec:simulations_and_models}

In this section we shall provide details about the propagation setup and the astrophysical model assumed for ultra-high energy cosmic ray sources.

\subsection{Propagation setup}
The sources were simulated as a 1D shell-like structure in the absence of extragalactic magnetic fields. The minimum distance considered was $D_{\min} = 1~\text{Mpc}$. Since expected neutrino flux from identical and isotropic nearby sources is negligible compared to that from more distant sources, varying the minimum distance does not change our results. Conversely, adopting an overly large maximum distance can lead to a significant overestimation of the expected neutrino fluxes \cite{alves_batista_2019_cosmogenic_photon_and_neutrino}. Since there is no definitive consensus, we adopted $z_{\max} \sim 6$, following recent studies on boosted neutrinos \cite{herrera_2025_diffuse_boosted_cosmic, herrera_2026_the_cosmic_neutrino_background_reach}. We simulated an equal distribution of the representative nuclear species $^1\text{H}$, $^4\text{He}$, $^{14}\text{N}$, $^{28}\text{Si}$, and $^{56}\text{Fe}$ in the energy range from $10^{18}~\text{eV}$ to $10^{21}~\text{eV}$. The energy losses included in the propagation were photopair production, photopion production, and photodisintegration, considering the CMB and the Gilmore's~et.~al.~\cite{gilmore_2012_semi_analytic_modelling_of_the_extragalactic} model for the EBL target photon fields. Adiabatic energy losses and nuclear decays were also taken into account.

For each simulation, we considered the sum of the two channels explained in Section \ref{sec:interactions_in_the_extragalactic_medium} for cosmic-ray C$\nu$B scattering, given an overdensity factor $\eta$ and a neutrino mass $m_\nu$. For the relic-neutrino target, we adopt a benchmark mass of $m_\nu = 0.1~\text{eV}$ while varying the overdensity factor from $\eta=10^2$ to $10^{10}$.

This value is chosen to facilitate comparison with previous boosted-\cnb studies. Since neutrino oscillations imply three distinct mass eigenstates, the physical \cnb should in principle be treated as a sum over eigenstates, each with its own mass and number density. A complete scan over neutrino mass ordering and individual mass eigenstates is beyond the scope of the present work. The behavior of the expected boosted \cnb component was previously explored in Ref. \cite{zhang_2026_impact_of_coherent} by varying the neutrino mass between $0.01~\text{eV}$ and $0.1~\text{eV}$, taking into account the same scattering channels used in the present work.

The present work neglects extragalactic magnetic fields and source intermittency. This approximation is appropriate for isolating the effect of competing photon- and neutrino-background interactions in a one-dimensional propagation setup, but it does not capture magnetic deflections, time delays, magnetic horizons, or the modification of the effective source distribution. Several studies have investigated UHECR propagation in extragalactic magnetic fields, including diffusion and magnetic-horizon effects, constrained simulations of the local Universe, and CRPropa-based propagation in magnetized large-scale-structure models \cite{alves_batista_2014_diffusion, hackstein_2018_simulations, Ding__2021, de_Oliveira_2022, de_Oliveira_2022_souza, Taylor_2023, Bister_2024, Abdul_Halim_2024_magnetic, shaw2025influencegalactichalouhecr,rossoni2025anisotropysignaluhecrsstructured}. These effects can modify the arrival directions and, in some regimes, the observed spectrum and composition of UHECRs. Since our main observable is the diffuse neutrino flux, and since neutrinos propagate rectilinearly after production, we defer a full three-dimensional magnetized treatment to future work.

\subsection{Modelling ultra-high-energy cosmic ray sources}

To evaluate the relative contribution of each channel to the observed neutrino flux and identify source conditions that maximize the relative boosted C$\nu$B contribution, we first consider two representative, though unrealistic, scenarios: pure-proton and pure-iron injection at the source. This choice is motivated by the significant impact of mass composition on the resulting neutrino flux, as explained in Section \ref{sec:interactions_in_the_extragalactic_medium}. Subsequently, to obtain consistent predictions for the expected neutrino flux, we adopt the astrophysical model and best-fit parameters provided by the Pierre Auger Observatory \cite{pierre_auger_2017_combined_fit_of_spectrum_and_composition}. Reference \cite{pierre_auger_collaboration_2023_constraining_the_sources_of_ultra} provides a more recent study of this kind that could also be considered. However, while its analysis is more sophisticated, the underlying model is more complex, and its assumptions regarding the astrophysical scenario are more restrictive. We opted for the combined fit from Ref. \cite{pierre_auger_2017_combined_fit_of_spectrum_and_composition} because it provides a more agnostic and general framework.

The model assumes that the differential injection rate of nuclear species $i$ at the sources can be written as
\begin{equation}
\mathcal{Q}_i(E,z) =
\mathcal{Q}_0\, f_i\, \psi(z)
\left(\frac{E}{10^{18}\,\mathrm{eV}}\right)^{-\gamma}
f_{\mathrm{cut}}(E,Z_i),
\end{equation}
\noindent where $\mathcal{Q}_i(E,z)$ is the comoving injection rate per unit energy, per unit time, and per unit volume. The parameter $\mathcal{Q}_0$ is therefore the overall source-emissivity normalization, not the flux measured at Earth. The quantity $f_i$ denotes the fraction of nuclei of species $i$ injected at a reference energy, normalized such that $\sum_i f_i=1$, $Z_i$ is the corresponding atomic number, $\gamma$ is the source spectral index, $R_{\mathrm{cut}}$ is the maximum rigidity, and $f_{\mathrm{cut}}(E,Z_i)$ describes the rigidity-dependent cutoff. The function $\psi(z)$ accounts for the cosmological evolution of the comoving source emissivity; in the absence of source evolution we set $\psi(z)=1$.

For the representative scenarios, we assume a pure-proton composition with $f_{\text{H}} = 1$ (and $f_{\text{He}} = f_{\text{N}} = f_{\text{Si}} = f_{\text{Fe}} = 0$) and a pure-iron composition with $f_{\text{Fe}} = 1$ (and $f_{\text{H}} = f_{\text{He}} = f_{\text{N}} = f_{\text{Si}} = 0$). The cut-off function is defined as a standard exponential cut-off,
\begin{equation}
f_{\text{cut}}(E, Z_i) = \exp\left(- \frac{E}{Z_i R_{\text{cut}}}\right),
\end{equation}
\noindent and we evaluate the results for various values of $R_{\text{cut}}$ and $\gamma$. For the representative scenarios, the normalization $\mathcal{Q}_0$ is chosen such that the propagated UHECR spectrum matches the Pierre Auger Observatory data at $E = 5 \times 10^{18}~\mathrm{eV}$.

For the Auger combined fit model, we adopt the best-fit parameters for the propagation model closest to ours, listed in Table 9 of Ref. \cite{pierre_auger_2017_combined_fit_of_spectrum_and_composition}. Specifically, we use the CTG model, which features a spectral index of $\gamma = 0.87$, a maximum rigidity of $R_{\text{cut}} = 10^{18.62}~\text{V}$, and mass fractions of $f_{\text{H}} = 0$, $f_{\text{He}} = 0$, $f_{\text{N}} = 88\%$, $f_{\text{Si}} = 12\%$, and $f_{\text{Fe}} = 0$. The cut-off function in this case is given by
\begin{equation}
f_{\text{cut}}(E, Z_i) = 
\begin{cases}
1, & E < Z_i R_{\text{cut}}, \\
\exp\left(1 - \frac{E}{Z_i R_{\text{cut}}}\right), & E \ge Z_i R_{\text{cut}}.
\end{cases}
\end{equation}
The source emissivity evolution is parameterized as $\psi(z)=(1+z)^m$, where the evolutionary index $m$ can vary depending on the class of astrophysical sources. For our baseline scenario, we assume no cosmological evolution ($m=0$), strictly following the combined fit model from the Pierre Auger Observatory. To illustrate the sensitivity of the neutrino flux to the source evolution, we perform a benchmark test by varying $m$ while keeping the injection parameters fixed. Although this approach technically breaks the Auger combined fit, the effect is negligible for our high-energy regime of interest. Because the ultra-high-energy cosmic-ray flux above the ankle ($E \ge 5 \times 10^{18}~\text{eV}$) is dominated by local sources ($z \lesssim 1$), increasing $m$ primarily enhances the contribution of high-redshift sources at energies \textit{below} the ankle. Consequently, varying $m$ does not significantly compromise the baseline astrophysical model from Ref.~\cite{pierre_auger_2017_combined_fit_of_spectrum_and_composition}, which is explicitly fitted only for energies above the ankle. The impact of this source evolution on the observed neutrino fluxes is discussed below.
\section{Results}
\label{sec:results}

\subsection{Pure-proton and pure-iron scenarios}

A key strength of our methodology lies in its ability to directly compare the contributions of the three neutrino production channels arising from the same primary particles injected at the source. 

\begin{figure}[htbp]
        \centering
        \begin{minipage}[b]{0.82\linewidth}
            \centering
            \includegraphics[width=0.9\textwidth]{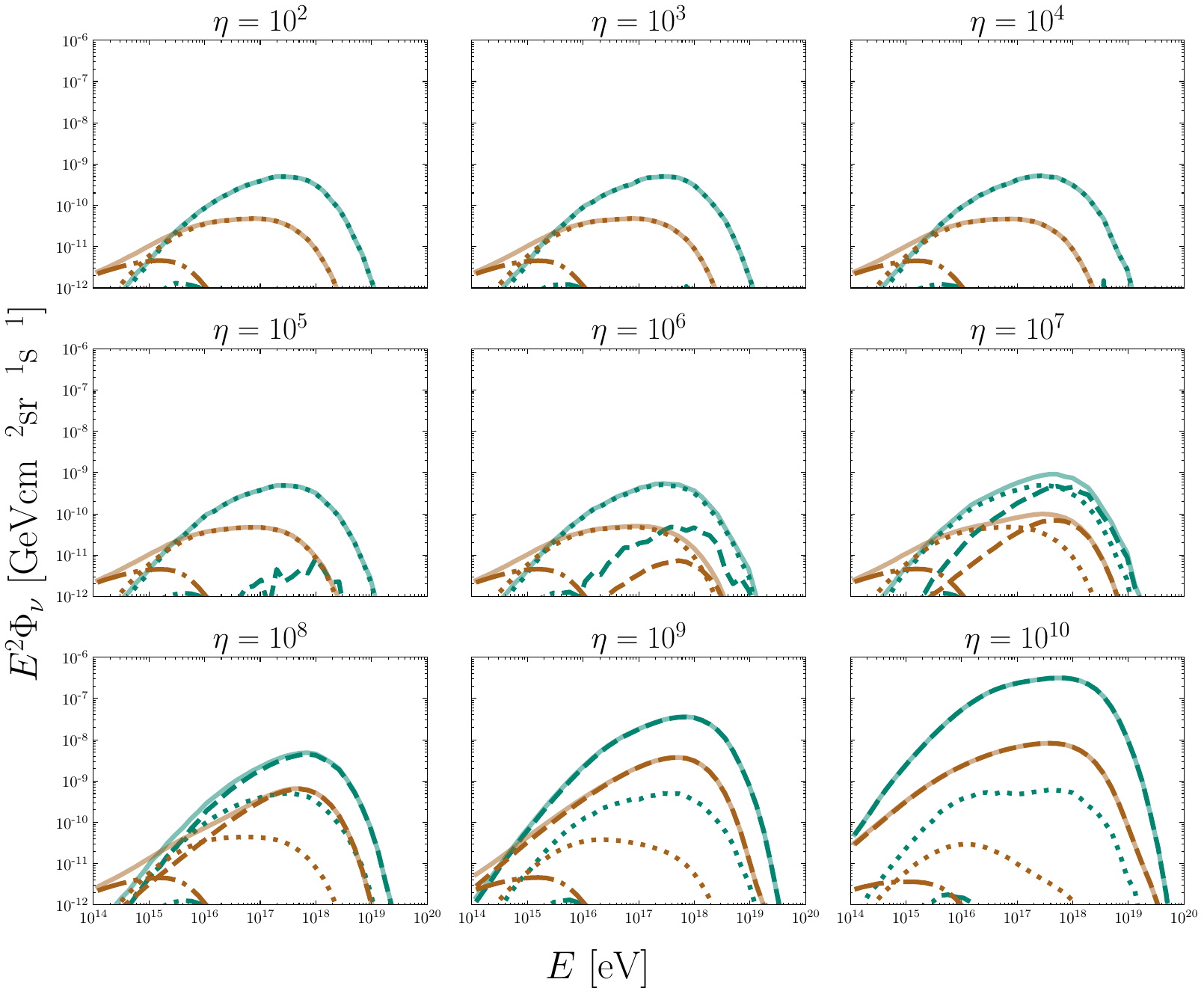}
        \end{minipage}
        \hspace{0.5cm}
        \begin{minipage}[b]{0.82\linewidth}
            \centering
            \includegraphics[width=0.9\textwidth]{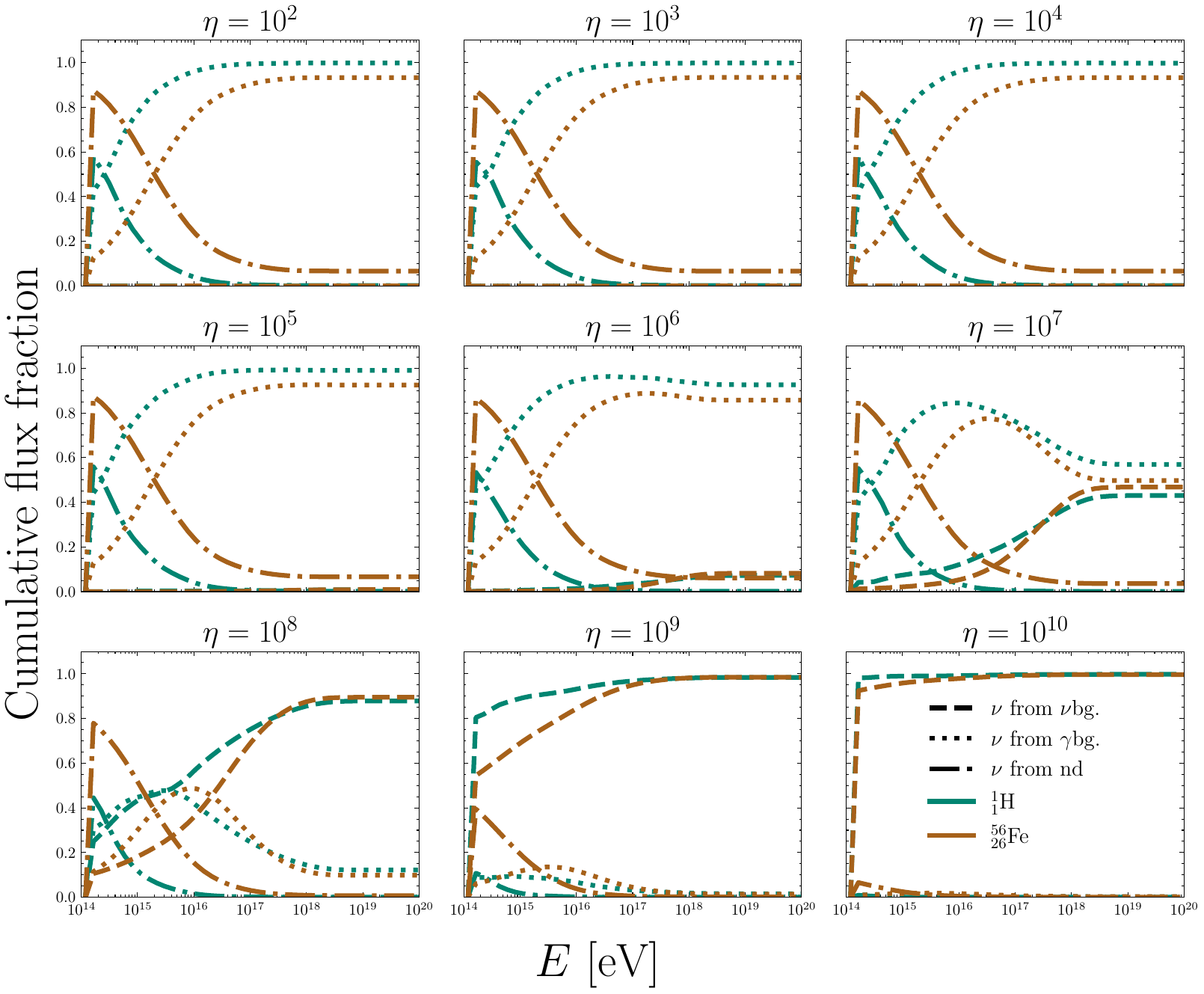}
        \end{minipage} 
        \caption{Expected differential neutrino fluxes ($E^2\Phi_\nu$, top panels) and cumulative flux fractions up to energy $E$ (bottom panels) for representative pure-proton ($^1_1\text{H}$) and pure-iron ($^{56}_{26}\text{Fe}$) injection scenarios. Colors denote the primary mass composition (e.g., orange for protons, cyan for iron). Line styles indicate the distinct neutrino production channels: solid lines represent the total neutrino flux, dashed lines represent boosted relic neutrinos, dotted lines represent neutrinos produced through $\gamma$BG interactions, and dash-dotted lines represent neutrinos from nuclear decay.}
        \label{fig:representative_scenario}
\end{figure}

The expected total neutrino flux at Earth and the fraction of the integrated flux for the representative pure-proton and pure-iron scenarios are shown in Fig.~\ref{fig:representative_scenario} for a range of different overdensity factors, assuming $R_\text{cut} = 10^{19}~\text{eV}$ and a source spectral index of $\gamma = 2$. The resulting fluxes reach magnitudes of $\sim 10^{-9}\text{--}10^{-7}~\text{GeV cm}^{-2}\text{sr}^{-1}\text{s}^{-1}$. However, these values should be treated as the upper and lower bounds for the expected flux, given that such pure mass compositions are highly unrealistic.

The panel arrangement in Fig.~\ref{fig:representative_scenario} clearly illustrates distinct physical regimes. In the top row ($\eta = 10^2\text{--}10^4$), interactions with the \photonbg completely dominate the total flux. For the pure-iron scenario, nuclear decays also provide a significant contribution to the low-energy tail of the spectrum. The middle row ($\eta = 10^5\text{--}10^7$) exhibits a clear transition regime, where the C$\nu$B contribution becomes increasingly competitive with the standard photon-background channels. At this point, it is crucial to address the impact of $R_\text{cut}$. For $R_\text{cut} = 10^{19}~\text{V}$, the C$\nu$B up-scattering component becomes comparable to the high-energy cosmogenic neutrino flux around $\eta \sim 10^7$. For lower rigidities, this transition overdensity tends to decrease. This occurs because the production of cosmogenic neutrinos via the \photonbg is heavily constrained by the maximum energy attained at the source, owing to the high energy threshold of the GZK processes. Consequently, the further the maximum rigidity falls below the GZK threshold, the lower the overdensity factor required for the boosted C$\nu$B component to become comparable to the cosmogenic neutrino background. The last row ($\eta = 10^8\text{--}10^{10}$) illustrates the regime in which the boosted C$\nu$B component becomes phenomenologically competitive with the photon-background and decay channels. Notably, at extremely large overdensity factors ($\eta \gtrsim 10^9$), the mean free path for C$\nu$B scattering becomes so short that it actively modifies the expected neutrino flux from the \photonbg.

The effects of the source spectral index, $\gamma$, and the cosmological evolution of the source emissivity, $\psi(z)$, primarily act as overall scaling factors for the expected neutrino flux. Generally, harder injection spectra (i.e., smaller values of $\gamma$) increase the fraction of high-energy particles at the source, thereby enhancing the resulting neutrino flux. Conversely, softer spectra yield the opposite effect. The cosmological evolution of the sources follows a similar trend, albeit with an important caveat. Because C$\nu$B scattering for low overdensity factors only becomes probable at high redshifts, source classes characterized by strong local emissivity and weak high-redshift evolution will yield a suppressed boosted C$\nu$B component. Consequently, in such scenarios, the effects of the \photonbg will even more strongly dominate over the up-scattered relic neutrinos.

\newpage

\subsection{Realistic scenario}

\begin{figure}[htbp]
    \centering
    \includegraphics[width=0.8\textwidth]{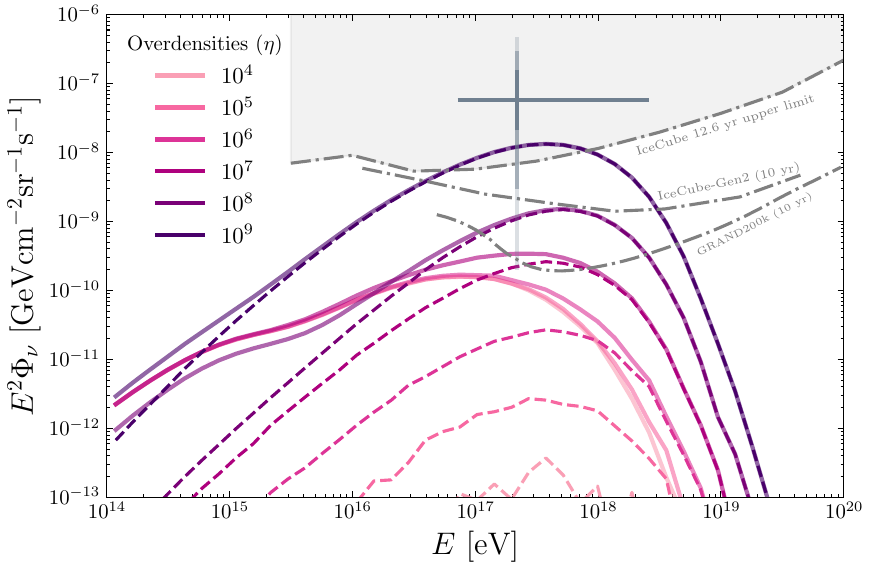}
    \caption{Expected diffuse neutrino flux ($E^2\Phi_\nu$) at Earth as a function of energy for varying C$\nu$B overdensity factors ($\eta$). Calculations are based on the combined fit astrophysical model from the Pierre Auger Observatory \cite{pierre_auger_2017_combined_fit_of_spectrum_and_composition}. The solid lines represent the total expected neutrino flux, while the dashed lines indicate the isolated contribution from up-scattered \cnb. For comparison, the current 12.6-year upper limit from IceCube \cite{icecube_2025_search_for_extremely} and the projected 10-year sensitivities for future observatories (IceCube-Gen2 and GRAND200k) \cite{icecube_2021_icecube_gen2_the_window,grand_2019_the_giant_radio_array} are shown in gray. The gray cross indicates the required flux to reproduce the KM3-230213A event detected by the KM3NeT Collaboration \cite{km3net_2025_observation} at the $90\%$, $95\%$, and $97.5\%$ confidence levels.}
    \label{fig:nu_flux_auger_2017}
\end{figure}

Fig.~\ref{fig:nu_flux_auger_2017} presents our results for the expected neutrino flux at various overdensities, calculated using the combined fit from the Pierre Auger Observatory. These results are shown alongside the projected sensitivities of future neutrino experiments, such as IceCube-Gen2 \cite{icecube_2021_icecube_gen2_the_window} and GRAND \cite{grand_2019_the_giant_radio_array}. We also include the current upper limit established by 12.6 years of IceCube observations \cite{icecube_2025_search_for_extremely}.

Note that the best-fit maximum rigidity of $R_\text{cut} = 10^{18.62}~\text{V}$ is lower than the $10^{19}~\text{V}$ value assumed in the representative scenarios of Fig.~\ref{fig:representative_scenario}. As a result, the contribution at the highest energies from C$\nu$B up-scattering becomes non-negligible at a lower overdensity factor of $\eta \sim 10^5$. Maximum rigidities of this order are not unique to this particular fit; low-rigidity or rigidity-limited source scenarios have also been discussed in phenomenological interpretations of Auger data and in source-population studies \cite{aloisio_2011_disappointing_model,ehlert_2023_maximum_rigidity_distribution,globus_2023_uhecr_source_models}. Nevertheless, in the present work we do not attempt to model a specific accelerator class. We use the Auger combined-fit value as a phenomenologically motivated benchmark for studying the relative importance of photon-background and C$\nu$B interactions.

For overdensity factors $\eta \gtrsim 10^7$, the expected neutrino flux falls within the 10-year projected sensitivity of GRAND200k, a regime where the dominant contribution is precisely the boosted C$\nu$B component. As the overdensity increases to $\eta \gtrsim 10^8$, the flux enters the projected sensitivity range of IceCube-Gen2 observatory. Assuming a neutrino mass of $m_\nu = 0.1~\text{eV}$, an extreme overdensity of $\eta = 10^9$ is already excluded by current IceCube limits within the context of this astrophysical model. For lower overdensities, the boosted C$\nu$B contribution remains subdominant to the neutrino flux produced by interactions with photon backgrounds.

\begin{figure}[htbp]
        \begin{minipage}[b]{0.49\linewidth}
            \centering
            \includegraphics[width=\textwidth]{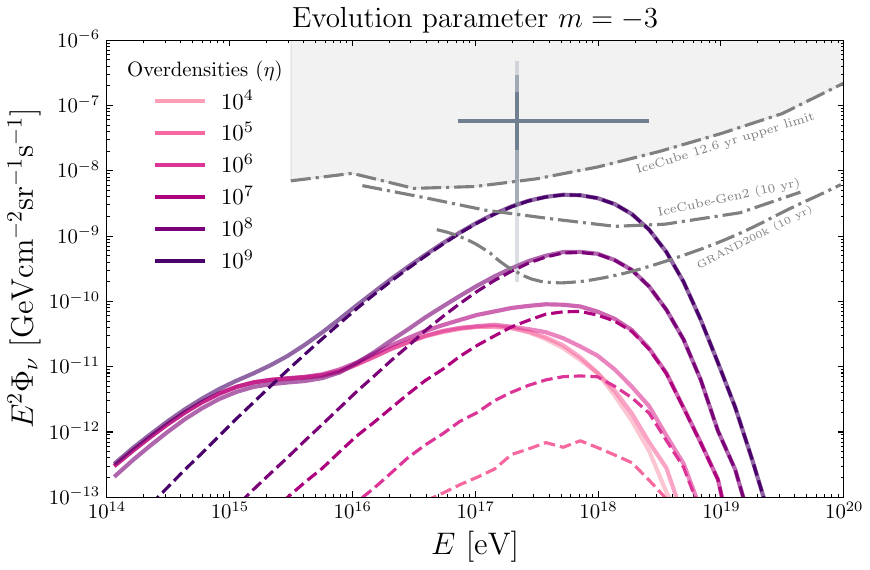}
        \end{minipage}
        \hspace{0.5cm}
        \begin{minipage}[b]{0.49\linewidth}
            \centering
            \includegraphics[width=\textwidth]{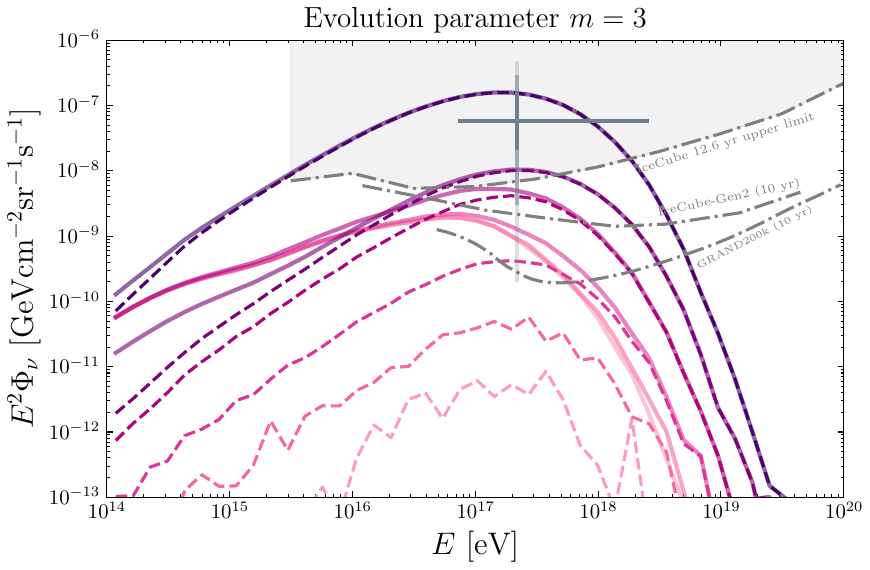}
        \end{minipage} 
    \caption{Expected neutrino fluxes at Earth, illustrating the sensitivity of the results to the cosmological source emissivity evolution. As a benchmark test, the evolution index ($m$) is varied while keeping the injection parameters fixed to the baseline combined fit model (the same underlying astrophysical model as in Fig.~\ref{fig:nu_flux_auger_2017}). The panels contrast two representative evolutionary scenarios: a negative evolution ($m = -3$, left panel) and a positive evolution ($m = 3$, right panel).}
        \label{fig:evolution}
\end{figure}

To illustrate the sensitivity of the expected neutrino flux to the cosmological evolution of the sources, Fig.~\ref{fig:evolution} presents our results incorporating an emissivity evolution function of the form $\psi(z) = (1 + z)^m$, specifically comparing the cases of $m = 3$ and $m = -3$.

\begin{figure}[htbp]
        \centering
        \includegraphics[width=0.75\textwidth]{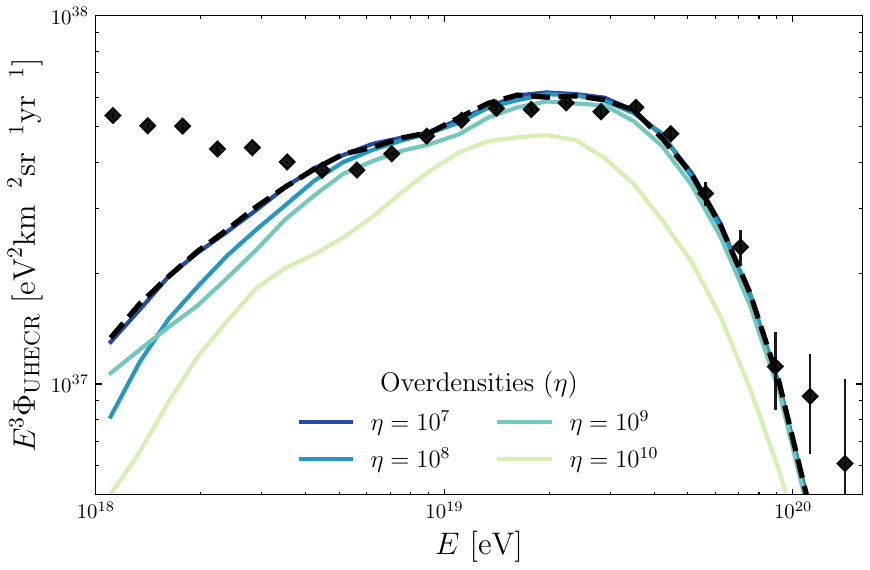}
        \caption{Ultra-high-energy cosmic ray energy spectrum at Earth illustrating the flux suppression induced by extreme \cnb overdensity factors ($\eta$). The spectra are calculated using the source parameters from the Pierre Auger Observatory combined fit model \cite{pierre_auger_2017_combined_fit_of_spectrum_and_composition}. The black diamonds represent the measured UHECR energy spectrum data from the Pierre Auger Observatory at epoch \cite{pierre_auger_2017_data}. The dashed black line represents the baseline UHECR flux, where C$\nu$B scattering was not included. Solid colored lines denote the attenuated fluxes for varying overdensities.} 
        \label{fig:uhecr_flux_auger_2017}
\end{figure}

In agreement with past studies \cite{demarchi_2025_relic_neutrino_background_cosmic_ray, zhang_2026_impact_of_coherent}, we find that extreme C$\nu$B overdensity values can significantly alter the expected ultra-high-energy cosmic ray spectrum at Earth. Because our Monte-Carlo approach inherently tracks all energy losses during propagation, it allows us to explicitly compute the impact of C$\nu$B scattering on the final UHECR spectrum, as shown in Fig.~\ref{fig:uhecr_flux_auger_2017}. 

Since the high-energy tail of the UHECR spectrum is predominantly governed by local sources ($z < 1$), the spectral modifications begin to manifest at energies just below the ankle. These suppressions become drastic when the rate of UHECR interactions with relic neutrinos becomes comparable to that of interactions with the \photonbg ($\eta \gtrsim 10^9$). It is worth noting, however, that such extreme overdensity factors are already excluded by the diffuse neutrino flux limits discussed. Quantitative thresholds on the overdensity factor remain subject to uncertainties associated with source evolution, EBL modeling, and nuclear interaction cross sections.

The results shown in Figs.~\ref{fig:representative_scenario}--\ref{fig:evolution} allow us to identify which aspects of the boosted-C$\nu$B signal are robust against the source assumptions explored here. In the representative pure-proton and pure-iron scenarios of Fig.~\ref{fig:representative_scenario}, photon-background interactions dominate the neutrino flux for small and moderate overdensities, while the boosted-C$\nu$B component becomes competitive only when the overdensity is increased to very large values. The same qualitative behavior is found in the Auger combined-fit scenario shown in Fig.~\ref{fig:nu_flux_auger_2017}, although the transition occurs at a somewhat lower overdensity because of the lower best-fit maximum rigidity. Finally, Fig.~\ref{fig:evolution} shows that changing the source emissivity evolution modifies the overall normalization of the predicted neutrino fluxes, but does not remove the need for large C$\nu$B overdensities for the boosted component to dominate over the standard photon-background and decay channels. We therefore interpret the suppression of the observable boosted-C$\nu$B contribution as a generic consequence of realistic UHECR propagation in the source scenarios considered here, rather than as a feature of a single benchmark model.
\section{Conclusion}
\label{sec:conclusion}

In this work, we reassessed previous expectations for boosted relic neutrino fluxes generated during ultra-high-energy cosmic ray propagation. Simplified approaches overestimate boosted relic neutrino fluxes because they do not fully account for realistic propagation losses and competing photon-background interactions. We improved upon previous analytical estimates by implementing a comprehensive Monte-Carlo framework using \crpropa. This approach allowed us to properly track the primary cosmic ray propagation, inherently accounting for energy losses, nuclear photodisintegration, and the unavoidable, competing interactions with diffuse photon backgrounds. By simulating the production of secondary neutrinos from standard photon-background interactions, nuclear decays, and neutral-current scattering on relic neutrinos, we provided a rigorous and consistent evaluation of the expected diffuse neutrino flux at Earth. 

As shown explicitly in Figs.~\ref{fig:representative_scenario}--\ref{fig:evolution}, the boosted relic-neutrino component remains subdominant to the standard photon-background and decay channels for small and moderate overdensities in all source scenarios explored here. Although the precise transition overdensity depends on the injected composition, maximum rigidity, and source evolution, the need for large C$\nu$B overdensities is a common feature of our propagation-based calculations. This indicates that realistic propagation effects, especially the depletion and reshaping of the UHECR population by photon-background interactions and continuous energy losses, are essential when estimating the observable boosted-C$\nu$B flux.

Our results demonstrate that the interactions with the \photonbg heavily suppress the expected flux of boosted \cnb neutrinos. Using the realistic mixed-composition astrophysical model derived from the Pierre Auger Observatory combined fit, we found that only substantial overdensities yield boosted C$\nu$B fluxes potentially accessible to next-generation observatories. Specifically, overdensity factors of $\eta \gtrsim 10^7\text{--}10^8$ are required for the up-scattered \cnb component to overcome the cosmogenic neutrino background and reach the projected 10-year sensitivities of next-generation observatories such as GRAND200k and IceCube-Gen2. Importantly, we showed that the boosted \cnb component and the standard \photonbg channels exhibit markedly different spectral shapes, providing a potentially useful spectral signature for future phenomenological discrimination studies.

Furthermore, our simulation explicitly revealed that only extreme overdensity factors significantly modify the primary UHECR spectrum. For $\eta \gtrsim 10^9$, the UHECR scattering rate with relic neutrinos becomes comparable to canonical \photonbg interactions, inducing a prominent suppression in the cosmic ray spectrum at energies just below the ankle. However, our coupled analysis demonstrates that such extreme values are already strongly disfavored within the context of the adopted astrophysical model by current IceCube upper limits. 

Our results indicate that the suppression of boosted relic neutrino fluxes emerges naturally from the hierarchy between photon-background and relic-neutrino interaction probabilities during realistic UHECR propagation. The suppression is not tied to a specific astrophysical scenario, but emerges naturally from the hierarchy between photon-background and relic-neutrino interaction probabilities during UHECR propagation.

Ultimately, while next-generation neutrino observatories may constrain extreme overdensities, accurately accounting for the dominant propagation effects is essential to avoid significantly overestimating the expected boosted relic neutrino signal. The dominance of photon-background interactions over relic-neutrino scattering emerges naturally from the relative interaction scales governing realistic UHECR propagation. 

Extensions of this work can be pursued. First, the calculation can be generalized by treating the three relic-neutrino mass eigenstates separately, including their individual masses, redshift evolution, and possible clustering in local gravitational potentials. Second, the present one-dimensional propagation setup can be extended to three-dimensional simulations including extragalactic magnetic fields, source discreteness, and source intermittency. Finally, the predicted boosted-C$\nu$B spectra should be folded with the detector response of upcoming and proposed high-energy neutrino observatories, including IceCube-Gen2, GRAND200k, POEMMA, RNO-G, PUEO, BEACON, Trinity, and TAMBO \cite{icecube_gen2_2021_window,grand_2020_science,poemma_2021_observatory,rnog_2021_design,pueo_2021_payload,beacon_2020_concept,trinity_2021_observatory,tambo_2025_deep_valley}. Such studies would allow the overdensity bounds derived here to be translated into experiment-specific discovery prospects and exclusion sensitivities.

\acknowledgments

The authors are supported by the S\~{a}o Paulo Research Foundation (FAPESP) through grant numbers 2021/01089-1, 2024/22722-2 and 2019/10151-2. VdS is supported by CNPq through grant number 308837/2023-1. The authors acknowledge the National Laboratory for Scientific Computing (LNCC/MCTI,  Brazil) for providing HPC resources for the SDumont supercomputer (http://sdumont.lncc.br).


\bibliographystyle{JHEP}
\bibliography{biblio.bib}

\end{document}